\documentclass[12pt]{article}

\usepackage{amsmath}
\usepackage{amssymb}
\usepackage{cite}

\def\nn{\nonumber}

\def\rmd{{\rm{d}}}

\numberwithin{equation}{section}

\title{\bf \Large Perfect fluid equations
with nonrelativistic\\ conformal supersymmetries}

\author{Timofei  Snegirev${}^{a}$\thanks{timofei.v.snegirev@tusur.ru}
\\[0.5cm]
\it{\small ${}^a$Laboratory of Applied Mathematics and Theoretical Physics,}\\
\it{\small Tomsk State University of Control Systems and Radioelectronics,}\\
\it{\small Lenin ave. 40, 634050 Tomsk, Russia}}

\date{}

\begin{document}

\maketitle

\begin{abstract}
Our recent result on the construction of perfect fluid equations
with $N=1,2$ Schr\"odinger supersymmetry [Mod. Phys. Lett. A 41 (2026) 2550214] is extended to accommodate
nonrelativistic conformal supersymmetries of other types. Two
cases are considered in detail, which include the $N=2$ conformal
Newton-Hooke superalgebra and $N=1$ $\ell$-conformal Galilei
superalgebra with arbitrary half-integer parameter $\ell$.
Supersymmetric
fluid models are built within the Hamiltonian framework by introducing
real (for $N=1$) or complex (for $N=2$) anticommuting field
variables as superpartners for the density and velocity. For both
the cases the full set of conserved charges associated with the
superalgebras is constructed and the Lagrangian description is
given. Subtleties with the construction of  perfect fluid equations with $N=2$
$\ell$-conformal Galilei supersymmetry are discussed as well.

\end{abstract}

\thispagestyle{empty}
\newpage
\setcounter{page}{1}

\allowdisplaybreaks

\section{Introduction}\label{Sec1}

Hydrodynamics provides universal effective description of
many-body systems at low frequencies and large wave-lengthes. In the
case of strongly-coupled systems, fluid models with conformal
symmetries play the crucial role. They are in the focus of the
fluid/gravity correspondence (for a review see \cite{Ran09}). At the
same time, the nonrelativistic AdS/CFT correspondence
\cite{Son08,BM08,NS2010}, which extends the holography to cover
strongly-coupled condensed matter systems, generates renewed interest
in fluid mechanics with nonrelativistic conformal (super)symmetries.

The key feature of nonrelativistic conformal symmetries is that the
temporal and spatial coordinates scale differently under the
dilatations: $t'=\lambda t$, $x'_i=\lambda^\ell x_i$. The anisotropic
scaling is characterized by the parameter $\ell$, the reciprocal of
which is known as the dynamical critical exponent. Although such
conformal symmetry is typically revealed in condensed matter systems,
it has been widely used in other physical contexts
\cite{DFF76,Hen02,CO03,Hor08,RCGV09,AG22}. For example, the anisotropic
scaling was used by Horava \cite{Hor08} to construct the model of
quantum gravity which is power-counting renormalizable and reduces
to the general relativity at large scales.

In addition to the dilatations, conformal extensions of the Galilei
algebra also include a special conformal transformation. In general,
the algebra is infinite-dimensional for an arbitrary parameter
$\ell$. The finite-dimensional extensions, which are the focus of
this paper, are possible for positive integer or half-integer $\ell$
only \cite{Hen97,NOR97}. They also include additional $2\ell-1$
vector generators associated with the accelerations. In literature,
such conformal extensions are usually referred to as the
$\ell$-conformal Galilei algebra. A particular instance of the
latter is the well-known Schr\"odinger algebra
\cite{Jack72,Nied72,Hag72,DHHR24} which is revealed for
$\ell=\frac12$. Note that the nonrelativistic contraction of the
relativistic conformal algebra reproduces the case of $\ell=1$.

Aiming at a better understanding of physical meaning of the
parameter $\ell$, dynamical realizations of the $\ell$-conformal
Galilei algebra have recently been studied (see e.g.
\cite{DH,FIL,DH1,GK,GM3,AGKM,AGGM,CG} and references therein).
Equations of motion for such systems typically include higher
derivative terms. To
give a notable example, the celebrated Pais-Uhlenbeck oscillator
\cite{PU} of order $(2\ell+1)$ enjoys the
$\ell$-conformal Galilei symmetry for a special choice of its
frequencies \cite{AGGM}.

Historically, the $N=1$ and $N=2$ $\ell=\frac12$ conformal
Galilei superalgebras were first revealed in the model of a
nonrelativistic spin-$\frac12$ particle \cite{GGT90} and in the
model of a harmonic oscillator extended by fermionic degrees of
freedom \cite{BH86}, respectively. Later, supersymmetric extensions of
$\ell=\frac12$ and $\ell=1$ conformal Galilei algebras were
extensively studied in various physical
contexts \cite{Horv93,DH94,HU05,AL09,Gal09,FL11}. Some of them
were further generalized to the case of an
arbitrary (half)-integer $\ell$
\cite{Mas11,Aiz12,AKT13,Mas14,GM17,GK17,MM19,GM21}. In particular,
$N=1,2,3,4$ $\ell$-conformal Galilei superalgebras were constructed.
It was shown that, in order to close the superalgebra for $N\geq 2$
and $\ell>\frac12$, it is necessary to include extra bosonic vector
generators.

It is well known that the perfect fluid equations described by the
continuity equation and the Euler equation enjoy the $\ell=\frac12$
conformal Galilei symmetries provided a polytropic equation of
state with a special exponent is imposed \cite{HH1,RS00,HZ}. Quite
recently, these equations have been generalized to accommodate the
$\ell$-conformal Galilei symmetries for an arbitrary value of $\ell$
\cite{Gal22a} (for related further developments see
\cite{Gal22b,Sne23a,Sne24,Sne25a}) . They include the generalized
Euler equation with higher-order material derivatives and the polytropic
equation of state with an exponent depending on $\ell$. Some exact
solutions for these generalized equations were found in
\cite{Gal26}, one of them is similar to the Bjorken flow \cite{Bj}.

Super-extended perfect fluid equations invariant under the
$\ell$-conformal Galilei supergroup have been studied for
$\ell=\frac12$ only \cite{Gal24,Sne25b}. The goal of this paper is
to generalize these results to the case of the $N=1$ $\ell$-conformal
Galilei supersymmetry with an arbitrary half-integer $\ell$, as well
as to the case of the $N=2$ conformal Newton-Hooke supersymmetries
\cite{Gal09}. The latter is the counterpart of the Schr\"odinger
superalgebra in the presence of a cosmological constant. Note that
supersymmetric extensions of nonrelativistic fluid mechanics are of
interest for several reasons. On the one hand, in $1+1$ and $1+2$
dimensions they are tightly connected with superstring and
supermembrane theories \cite{JP00}. On the other hand, some of them
provide nontrivial examples of integrable field theories
\cite{DP02}.

Like in our previous publication \cite{Sne25b}, we choose
to work within the Hamiltonian
framework. Supersymmetric construction described below amounts to
introducing fermionic field partners for the bosonic degrees of
freedom and building a single real (for $N=1$) or complex (for
$N=2$) supercharge, which via the Poisson bracket generates the
super-extended Hamiltonian. The latter governs the dynamics of the
enlarged system and reduces to the original theory in the
bosonic limit. Note that this scheme does not guarantee that all the
desired superconformal symmetries automatically hold. As shown below,
the construction of perfect fluid equations with the $N=2$ $\ell$-conformal Galilei
supersymmetries presents a challenge.

The work is organized as follows.  In the next section, we briefly
review the perfect fluid equations with $N=1,2$ Schr\"odinger
supersymmetry within the Hamiltonian framework \cite{Sne25b}. In
Sect. \ref{Sec3}, these results are extended to include a
cosmological constant. The perfect fluid equations with the $N=2$
conformal Newton-Hooke supersymmetry are constructed. In Sect.
\ref{Sec4}, by introducing a set of fermionic fluid variables, a
generalized perfect fluid model is built, which is invariant under
the $N=1$ $\ell$-conformal Galilei superalgebra. The possibility to
use a similar consideration to construct generalized perfect fluid
equations with $N=2$ $\ell$-conformal Galilei supersymmetry is
discussed. In Sect. \ref{Sec5}, we summarize our results and discuss
possible further developments. Generalized $N=2$ supersymmetric
fluid equations which are used in Sect. \ref{Sec4} are gathered in
Appendix.

\section{Perfect fluid equations with the Schr\"odinger supersymmetry}\label{Sec2}

We begin by a brief account of the perfect fluid equations with
$N=1,2$ Schr\"odinger supersymmetries (for more details see
\cite{Sne25b}). On the one hand, this allows us to establish
general features of the supersymmetric construction to be used below. On
the other hand, the results in this section are the limiting case
of those presented in the following sections.

\subsection{Perfect fluid equations}

In non-relativistic space-time parameterized by the coordinates
$(t,x_i)$, $i=1,...,d$, perfect fluid equations are given
by\footnote{Throughout the text, we use the notations:
$\partial_0=\frac{\partial}{\partial t}$,
$\partial_i=\frac{\partial}{\partial x_i}$, ${\cal
D}=\partial_0+\upsilon_i\partial_i$ where $\upsilon_i$ is the fluid
velocity vector. Summation over repeated indices is understood.
Considering the coordinates $t$ and $x_i$ as independent, one has
the identity ${\cal D}x_i=\upsilon_i$.} \cite{LL87}
\begin{eqnarray}\label{PFEq}
{\partial_0\rho}+ {\partial_i (\rho\upsilon_i)}=0,\quad \rho{\cal
D}\upsilon_i=-{\partial_i p}+{f_i},
\end{eqnarray}
where the first one is the continuity equation for density
$\rho(t,x)$ and the second one is the Euler equation to determine $\upsilon_i(t,x)$.
Here,
$\upsilon_i(t,x)$ is the fluid velocity vector,  ${\cal
D}=\partial_0+\upsilon_i\partial_i$ is the material derivative,
$p(t,x)$ is the pressure which is assumed to be related to the
density via an equation of state $p=p(\rho)$ and $f_i\equiv{\bf f}$
designate external forces. For example, the latter can be ${\bf
f}=\rho{\bf g}$ for the gravitational force with acceleration ${\bf g}$
or the Lorentz force ${\bf f}=\frac{1}{c}{\bf j}\times{\bf H}$ due to the
magnetic field ${\bf H}$.

For $f_i=0$, the equations of motion (\ref{PFEq}) can be obtained
from the Hamiltonian
\begin{eqnarray}\label{Ham}
H=\int \rmd
x\left(\frac12\rho\upsilon^{}_i\upsilon^{}_i+V\right),\quad \rmd
x=\rmd x_1\rmd x_2...\rmd x_d,
\end{eqnarray}
provided the non-canonical Poisson brackets are introduced
\cite{MG80}
\begin{eqnarray}\label{PBPF}
\{\rho(x),\upsilon_i(x')\}=- {\partial_i}\delta(x,x'), \quad
\{\upsilon_i(x),\upsilon_j(x')\}=\frac{1}{\rho}\omega_{ij}\delta(x,x'),
\end{eqnarray}
where $\omega_{ij}={\partial_i\upsilon_j} -{\partial_j\upsilon_i}$
is the fluid vorticity. The potential function $V(\rho)$ in
(\ref{Ham}) links to the pressure via the Legendre transformation
$p=\rho V'-V$ (prime denotes differentiation with respect to the
argument).

In what follows, we assume that the fluid pressure obeys the
polytropic equation of state $p=\lambda(\gamma-1)\rho^\gamma$, where
$\lambda>0$ is a constant and $\gamma\neq0,1$ is the polytropic
exponent. It yields the potential
\begin{eqnarray}\label{PolPot}
V=\frac{1}{\gamma-1}p=\lambda\rho^\gamma,
\end{eqnarray}
which enters the Hamiltonian (\ref{Ham}).

It is well known that for the polytropic exponent
$\gamma=1+\frac{2}{d}$, where $d$ is the spatial dimension, the
perfect fluid equations (\ref{PFEq}) hold invariant under the
Schr\"odinger group.  Apart from the conventional time translations,
space translations, space rotations and Galilei boosts, it involves
the dilatations and  special conformal transformations. Denoting the
corresponding conserved charges by $H$, $P_i$, $M_{ij}$, $C_i$, $D$
and $K$, respectively, one has
\begin{eqnarray}\label{SrCons}
&&P_i^{}=\int \rmd x \rho \upsilon^{}_i,\quad C_i^{}=tP_i-\int \rmd
x \rho x_i,\nn
\\
&& M_{ij}=\int \rmd x (\rho\upsilon_ix_j-\rho\upsilon_jx_i),\quad
D=tH-\frac12\int \rmd x\rho\upsilon_i^{}x_i,\nn
\\
&& K= -t^2H+2tD+\frac12\int \rmd x\rho x_ix_i,\quad m=\int \rmd
x\rho,
\end{eqnarray}
where $H$ is Hamiltonian (\ref{Ham}) and $m$ is the conserved total
mass which plays the role of the central charge. Under the Poisson
bracket, the integrals of motion (\ref{SrCons}) form a
representation of the Schr\"odinger algebra \cite{Nied72}
\begin{align}\label{ShrAl}
& {[H,D]}=H,  && [H,C^{}_i]=P_i,\nn
\\
& {[H,K]}=2D,  && [P_i^{},C_j^{}]=\delta_{ij} m,\nn
\\
& {[D,K]}=K, &&
[P_i,R_{jk}^{}]=\delta_{ij}P_{k}-\delta_{ik}P_{j},\nn
\\
& {{[D,P^{}_i]}=-\frac12P^{}_i}, &&
[C_i,M_{jk}^{}]=\delta_{ij}C_{k}-\delta_{ik}C_{j},\nn
\\
& {{[D,C^{}_i]}=\frac12C^{}_i}, &&
[M_{ij}^{},M_{kl}^{}]=\delta_{i[k}M_{l]j}-\delta_{j[k}M_{l]i},\nn
\\
& {{[K,P^{}_i]}=-C^{}_i},
\end{align}
and generate the corresponding symmetry transformations for $\rho$
and $\upsilon_i$.

\subsection{$N=1$ supersymmetry}

An $N=1$ supersymmetric extension of the Schr\"odinger algebra
\cite{GGT90} involves additional Grassmann-odd generators which
include the supercharge $Q$, the generator of superconformal
transformation $S$, and the superpartners $\Lambda_i, i=1,...,d$ of
the Galilei boosts. All together they obey the following
(anti)-commutation relations \cite{Horv93}
\begin{align}\label{ShrAlN1}
&  \{Q,Q\}=2iH, && \{Q,\Lambda_i\}=iP_i && [H,S]=Q,\nn
\\
& \{Q,S\}=2iD, && \{S,\Lambda_i\}=iC_i, && [D,S]=\frac12S,\nn
\\
& \{S,S\}=2iK, && \{\Lambda_i,\Lambda_j\}=i\delta_{ij}m, &&
[D,Q]=-\frac12Q,\nn
\\
& [K,Q]=-S, && [Q,C_i]=\Lambda_i, && [S,P_i]=-\Lambda_i,\nn
\\
&
[\Lambda_i,M_{jk}^{}]=\delta_{ij}\Lambda_{k}-\delta_{ik}\Lambda_{j}.
&&
\end{align}
In particular, it contains $osp(1|1)$ spanned by $H,D,K$ and
$Q,S$.

Following the approach developed within the Hamiltonian formalism
\cite{Sne25b}, an extended perfect fluid dynamics with the $N=1$
Schr\"odinger supersymmetries\footnote{For a superfield
hydrodynamics with the $N=1$ Schr\"odinger supersymmetry see
\cite{Gal24}.} is realized by introducing real Grassmann-odd
partners\footnote{We use the conventional notation for complex
conjugation of the Grassmann-odd variables $\psi_1$ and $\psi_2$:
$(\psi_1\psi_2)^*=\psi_2^*\psi_1^*$. In case of real fermions
$\psi^*=\psi$, one has $(\psi_1\psi_2)^*=-\psi_1\psi_2$. }
$\sigma(t,x)$ and $\psi_i(t,x), i=1,...,d$, for the Grassmann-even
fields characterizing the density $\rho(t,x)$ and the velocity
$\upsilon_i(t,x)$ and imposing the following Poisson brackets
compatible with the super-Jacobi identities:
\begin{align}\label{SrPB}
&\{\psi_i(x),\psi_j(x')\}=\frac{i}{\rho}\delta_{ij}\delta(x,x'), &&
\{\sigma(x),\sigma(x')\}=\frac{i}{\rho}\delta(x,x'),\nn
\\
&\{\upsilon_i(x),\psi_j(x')\}=\frac{1}{\rho}{\partial_i\psi_j}\delta(x,x'),
&&
\{\upsilon_i(x),\sigma(x')\}=\frac{1}{\rho}\partial_i\sigma\delta(x,x').
\end{align}

Given (\ref{SrPB}), the $N=1$ supersymmetry charge can be chosen in
the simplest form, which is linear in the Grassmann-odd variables
\begin{eqnarray}\label{N1SrCh}
Q=\int \rmd x
(\rho\upsilon_i\psi_i+\sqrt{2\lambda}\rho^{\frac{\gamma+1}{2}}\sigma).
\end{eqnarray}
Via the Poisson bracket, it gives rise to the super-extended
Hamiltonian
\begin{eqnarray}\label{HamN1}
\{Q,Q\}=i\int \rmd x
\Big(\rho\upsilon_i\upsilon_i+2\lambda\rho^{\gamma}-i\sqrt{2\lambda}(\gamma-1)\rho^\frac{\gamma+1}{2}\partial_i\psi_i\sigma\Big)=2iH.
\end{eqnarray}
The first two terms entering the last expression reproduce the
original bosonic theory (\ref{Ham}), while the last contribution
describes the boson-fermion coupling. From the super-Jacobi
identities it follows that the supercharge is conserved,
$\{Q,H\}=0$.

The super-extended Hamiltonian (\ref{HamN1}) generates the $N=1$
supersymmetric dynamics characterized by the following equations of
motion
\begin{eqnarray}\label{N1eq}
\partial_0\rho&=&\{\rho,H\}=-\partial_i(\rho\upsilon_i),\nn
\\
\partial_0\sigma&=&\{\sigma,H\}=-\upsilon_i\partial_i\sigma-\frac{\sqrt{2\lambda}}{2}(\gamma-1)\rho^{\frac{\gamma-1}{2}}\partial_i\psi_i,\nn
\\
\partial_0\psi_i&=&\{\psi_i,H\}=-\upsilon_j\partial_j\psi_i-\frac{\sqrt{2\lambda}}{2}(\gamma-1)\frac{1}{\rho}\partial_i(\rho^{\frac{\gamma+1}{2}}\sigma),\nn
\\
\partial_0\upsilon_i&=&\{\upsilon_i,H\}=-\upsilon_j\partial_j\upsilon_i-\lambda(\gamma-1)\frac{1}{\rho}\partial_i(\rho^{\gamma})+
\frac{1}{\rho}\partial_j\Sigma_{ij},
\end{eqnarray}
where we denoted
\begin{eqnarray}
\Sigma_{ij}=i\frac{\sqrt{2\lambda}}{4}(\gamma-1)\Big[(\gamma-1)\delta_{ij}(\rho^{\frac{\gamma+1}{2}}\partial_k\psi_k\sigma)
+2(\rho^{\frac{\gamma+1}{2}}\partial_i\psi_j\sigma)\Big].
\end{eqnarray}

The Grassmann-even part of the $N=1$ Schr\"odinger superalgebra is
realized by the integrals of motion (\ref{SrCons}) where $H$ is
replaced by the superextended Hamiltonian (\ref{HamN1}) and  $M_{ij}$
should be modified as follows
\begin{eqnarray}
M_{ij}&=&\int \rmd x (\rho\upsilon_ix_j-\rho\upsilon_jx_i)-i\int
\rmd x \rho\psi_i\psi_j.
\end{eqnarray}

The integrals of motion associated with the Grassmann-odd part read
\begin{eqnarray}
\Lambda_i=\int \rmd x \rho\psi_i,\quad S=Qt-\int \rmd
x\rho\psi_ix_i,
\end{eqnarray}
where $Q$ is defined by (\ref{N1SrCh}) and $S$ is time-independent
for $\gamma=1+\frac{2}{d}$ only.

Under the Poisson brackets the complete set of the integrals of motion
obey the structure relations of the $N=2$ Schr\"odinger superalgebra
(\ref{ShrAlN1}) with the central charge $m=\int \rmd x\rho$.

\subsection{$N=2$ supersymmetry}

An $N=2$ supersymmetric extention of the Schr\"odinger algebra
\cite{BH86,DH94} includes pairs of Grassmann-odd generators
describing the supercharges $Q,\bar{Q}$, the generators of
superconformal transformations $S,\bar S$, the superpartners
$\Lambda_i,{\bar\Lambda}_i$ for the Galilei boosts, as well as extra
Grassmann-even generator $J$, which corresponds to $u(1)$
$R$-symmetry (for the structure relations see \cite{Sne25b}).

In order to construct an $N=2$ supersymmetric extension of the
perfect fluid equations, we introduce complex anticommuting fields
$(\sigma,\psi_i)$ as superpartners for the density and velocity
$(\rho,\upsilon_i)$ and impose the following Poisson brackets
\begin{align}\label{PBN2}
&\{\psi_i(x),\bar\psi_j(x')\}=\frac{i}{\rho}\delta_{ij}\delta(x,x'),
&&\{\upsilon_i(x),\psi_j(x')\}=\frac{1}{\rho}{\partial_i\psi_j}\delta(x,x'),\nn
\\
& \{\sigma(x),\bar\sigma(x')\}=\frac{i}{\rho}\delta(x,x'), &&
\{\upsilon_i(x),\sigma(x')\}=\frac{1}{\rho}\partial_i\sigma\delta(x,x'),\nn
\\
&\{\psi_i(x),\psi_j(x')\}=0,
&&\{\upsilon_i(x),\bar\psi_j(x')\}=\frac{1}{\rho}{\partial_i\bar\psi_j}\delta(x,x'),\nn
\\
& \{\sigma(x),\sigma(x')\}=0, &&
\{\upsilon_i(x),\bar\sigma(x')\}=\frac{1}{\rho}\partial_i\bar\sigma\delta(x,x'),
\end{align}
where $(\bar\sigma,\bar\psi_i)$ are complex conjugates of
$(\sigma,\psi_i)$. It is straightforward to verify that the
super-Jacobi identities are satisfied.

As the next step, two self-adjoint supercharge generators are
constructed
\begin{eqnarray}\label{SupN2}
Q&=&\int \rmd x(
\rho\upsilon_i\psi_i+\sqrt{2\lambda}\rho^{\frac{\gamma+1}{2}}\sigma+\frac{i}{2}(\gamma-1)\rho\partial_i\psi_i\sigma\bar\sigma),\nn
\\
\bar Q&=&\int \rmd x
(\rho\upsilon_i\bar\psi_i+\sqrt{2\lambda}\rho^{\frac{\gamma+1}{2}}\bar\sigma-\frac{i}{2}(\gamma-1)\rho\partial_i\bar\psi_i\sigma\bar\sigma),
\end{eqnarray}
which via $\{Q,\bar Q\}=2iH$ give rise to the $N=2$ super-extended
Hamiltonian
\begin{eqnarray}\label{HamN2}
H&=&\int \rmd x
\Big(\frac12\rho\upsilon_i\upsilon_i+\lambda\rho^{\gamma}-i\frac{\sqrt{2\lambda}}{2}(\gamma-1)\rho^\frac{\gamma+1}{2}(\partial_i\psi_i\bar\sigma+\partial_i\bar\psi_i\sigma)\nn\\
&&\qquad\qquad
-\frac12(\gamma-1)\rho\partial_i\psi_j\partial_j\bar\psi_i\sigma\bar\sigma
+\frac{1}{8}(\gamma-1)^2\frac{1}{\rho}\partial_i(\rho\sigma\bar\sigma)\partial_i(\rho\sigma\bar\sigma)\Big).
\end{eqnarray}
In contrast to the $N=1$ case the supercharges involve terms cubic
in the fermionic fields which are required by the brackets $\{Q,Q\}=0$ and
$\{\bar Q,\bar Q\}=0$.

From the super-extended Hamiltonian one obtains the equations of
motion
\begin{eqnarray}\label{N2eq}
{\cal D}\rho&=&-\rho\partial_i\upsilon_i,\qquad \rho{\cal
D}\upsilon_i=-(\gamma-1)\partial_i(\lambda\rho^{\gamma})+\partial_j\Sigma_{ij}\nn
\\
{\cal
D}\sigma&=&-\frac{\sqrt{2\lambda}}{2}(\gamma-1)\rho^{\frac{\gamma-1}{2}}\partial_i\psi_i
+i\frac{\gamma-1}{2}\partial_i\psi_j\partial_j\bar\psi_i\sigma
+\frac{i}{4}(\gamma-1)^2\sigma\partial_i(\frac{1}{\rho}\partial_i(\rho\sigma\bar\sigma)),\nn
\\
\rho{\cal
D}\psi_i&=&-\frac{\sqrt{2\lambda}}{2}(\gamma-1)\partial_i(\rho^{\frac{\gamma+1}{2}}\sigma)
-i\frac{(\gamma-1)}{2}\partial_j(\rho\partial_i\psi_j\sigma\bar\sigma),
\end{eqnarray}
where we denoted
\begin{eqnarray}\label{SigTen}
\Sigma_{ij}&=&\frac12(\gamma-1)\Big[
i\frac{\sqrt{2\lambda}}{2}(\gamma-1)\delta_{ij}\rho^{\frac{\gamma+1}{2}}(\partial_k\bar\psi_k\sigma
+\partial_k\psi_k\bar\sigma)
+i\sqrt{2\lambda}\rho^{\frac{\gamma+1}{2}}(\partial_i\bar\psi_j\sigma+\partial_i\psi_j\bar\sigma)\nn
\\
&&\qquad\qquad\quad +\rho(\partial_i\psi_k\partial_k\bar\psi_j -
\partial_i\bar\psi_k\partial_k\psi_j)\sigma\bar\sigma
+\frac{(\gamma-1)}{2}\rho\sigma\bar\sigma\partial_i(\frac{1}{\rho}\partial_j(\rho\sigma\bar\sigma))\Big].
\end{eqnarray}
The equations of motion for $\bar\sigma$ and $\bar\psi_i$ can be
obtained from the equations for $\sigma$ and $\psi_i$ by the complex
conjugation.

The integrals of motion realizing the Grassmann-even part of the
$N=2$ Schr\"odinger superalgebra read
\begin{eqnarray}\label{N2IM_ev}
P_i^{}&=&\int \rmd x \rho \upsilon^{}_i,\qquad C_i^{}=tP_i-\int \rmd
x \rho x_i,\qquad J=\int \rmd x
\rho(\psi_i\bar\psi_i+\sigma\bar\sigma)\nn
\\
M_{ij}&=&\int \rmd x (\rho\upsilon_ix_j-\rho\upsilon_jx_i)-i\int
\rmd x \rho(\psi_i\bar\psi_j-\psi_j\bar\psi_i),\nn
\\
D&=&tH-\frac12\int \rmd x\rho\upsilon_i^{}x_i,\qquad K=
-t^2H+2tD+\frac12\int \rmd x\rho x_ix_i,
\end{eqnarray}
where $H$ is super-extended Hamiltonian (\ref{HamN2}). The last
two integrals of motion are present for a particular value of the
polytropic exponent $\gamma=1+\frac{2}{d}$ only. Note that, in
contrast to the $N=1$ case, there is an additional Grassmann-even
integral of motion $J$ which builds upon the quadratic
combinations of Grassmann-odd fields.

The Grassmann-odd part of the $N=2$ Schr\"odinger superalgebra is
realized by the supercharges $Q$ and $\bar Q$, as well as by the
integrals of motion
\begin{align}\label{N2IM_odd}
& \Lambda_i=\int \rmd x \rho\psi_i, && \bar\Lambda_i=\int \rmd x
\rho\bar\psi_i,\nn
\\
& S=Qt-\int \rmd x\rho\psi_ix_i, && \bar S=\bar Qt-\int \rmd
x\rho\bar\psi_ix_i,
\end{align}
where $S,\bar{S}$ correspond to the superconformal generators and
$\Lambda_i,\bar\Lambda_i$ represent the superpartners of the
Galilei boosts.

It is straightforward to verify that the conserved charges
(\ref{SupN2}),(\ref{HamN2}), (\ref{N2IM_ev}), and (\ref{N2IM_odd})
 obey the structure relations of the
$N=2$ Schr\"odinger superalgebra (for more detail see
\cite{Sne25b}) with the central charges $m=\int \rmd x\rho$ and $Z=0$.

\section{Perfect fluid equations with $N=2$ conformal Newton-Hooke supersymmetry}\label{Sec3}

It is well known that in the presence of a cosmological constant the
Schr\"odinger algebra is replaced by a conformal Newton-Hooke
algebra \cite{NOR97,Gal09,Gal10}. The former follows from the latter in
the limit when a cosmological constant tends to zero. The
corresponding structure relations take the form (\ref{ShrAl}) with
the replacements
\begin{align}\label{NHal}
& {[H,D]}=H\mp\frac{2}{R^2}K,  && {[H,P_i]}=\mp\frac{1}{R^2}C_i,
\end{align}
where $R$ is a characteristic time and the upper/lower sign corresponds to the negative/positive cosmological
constant $\Lambda\sim\mp\frac{1}{R^2}$. In the limit when
the cosmological constant tends to zero $\Lambda\rightarrow0$
($R\rightarrow\infty$) the Schr\"odinger algebra (\ref{ShrAl}) is
reproduced.

In a recent paper \cite{Sne25a}, it was shown that the perfect
fluid dynamics with the conformal Newton-Hooke symmetries\footnote{In
what follows, we consider the case of a negative cosmological constant.
For a positive cosmological constant a conventual supersymmetric
extension turns out to be problematic. It links to the difficulty
to define positive conserved energy in the parent de Sitter space.}
is described by the equations (\ref{PFEq}), in which the force term
reads $f_i=-\frac{1}{R^2}\rho x_i$ and the pressure obeys the
same polytropic equation of state as in the Schr\"odinger case
$p=\lambda(\gamma-1)\rho^\gamma$ with $\gamma=1+\frac{2}{d}$. The
Euler equation in (\ref{PFEq}) takes the form
\begin{eqnarray}\label{PFEqNH} {\cal
D}\upsilon_i+\frac{1}{R^2}x_i=-\frac{1}{\rho}{\partial_i p}.
\end{eqnarray}

Together with the continuity equation, (\ref{PFEqNH})
can be represented in the Hamiltonian form
\begin{eqnarray}
\partial_0\rho=\{\rho,H\}=-\partial_i(\rho\upsilon_i),\quad
\partial_0\upsilon_i=\{\upsilon_i,H\}=-\upsilon_j\partial_j\upsilon_i-\frac{1}{R^2}x_i-\frac{1}{\rho}{\partial_i p}
\end{eqnarray}
where
\begin{eqnarray}\label{NHHam}
H=\int \rmd
x\left(\frac12\rho\upsilon^{}_i\upsilon^{}_i+\frac{1}{2R^2} \rho
x_ix_i+V\right),\quad p=\rho V'-V,
\end{eqnarray}
and the Poisson brackets are exposed in (\ref{PBPF}).

The integrals of motion associated with the conformal Newton-Hooke
symmetries read
\begin{eqnarray}\label{NHCons}
P_i^{}&=&\int \rmd x (\rho
\upsilon^{}_i\cos{\frac{t}{R}}+\frac{1}{R}\rho x_i\sin{\frac{t}{R}}
),\quad M_{ij}=\int \rmd x (\rho\upsilon_ix_j-\rho\upsilon_jx_i),\nn
\\
C_i^{}&=&R\int\rmd x (\rho
\upsilon^{}_i\sin{\frac{t}{R}}-\frac{1}{R}\rho x_i\cos{\frac{t}{R}}
),\quad m=\int \rmd x \rho\nn
\\
D&=&\frac{R}{2}H\sin{\frac{2t}{R}}-\frac{1}{2}\int \rmd
x(\rho\upsilon_ix_i\cos{\frac{2t}{R}}+\frac{1}{R}\rho
x_ix_i\sin{\frac{2t}{R}})\nn
\\
K&=&\frac{R^2}{2}H(1- \cos{\frac{2t}{R}})-\frac{R}{2}\int \rmd
x(\rho\upsilon_ix_i\sin{\frac{2t}{R}}-\frac{1}{R}\rho
x_ix_i\cos{\frac{2t}{R}}).
\end{eqnarray}
Under the Poisson brackets, they satisfy the structure relations
(\ref{ShrAl}), (\ref{NHal}) and transit smoothly to (\ref{SrCons})
in the limit $R\rightarrow\infty$.

Let us turn to the construction of a supersymmetric model. Like in the
previous section, we extend the original Grassmann-even phase space
$(\rho,\upsilon_i)$ by introducing Grassmann-odd field variables
$(\sigma,\psi_i)$. It seems natural to impose Poisson brackets for
them as in the Schr\"odinger case, which are specified by
(\ref{SrPB}) and (\ref{PBN2}) for $N=1$ and $N=2$ supersymmetry,
respectively. As the next step, we need to construct a real (for
$N=1$) or complex (for $N=2$) supercharge that, via the Poisson
bracket, will define the super-extended Hamiltonian. One can show
that due to the presence of the harmonic potential term in the
Hamiltonian (\ref{NHHam}), this cannot be realized for $N=1$
supersymmetry. However, this can be done for the $N=2$ case .

Complex conjugate supercharges are derived from (\ref{SupN2}) with a
correction responsible for the cosmological term
\begin{eqnarray}
Q&=&\int \rmd x(
\rho\upsilon_i\psi_i+\sqrt{2\lambda}\rho^{\frac{\gamma+1}{2}}\sigma+\frac{i}{2}(\gamma-1)\rho\partial_i\psi_i\sigma\bar\sigma+\frac{i}{R}\rho\psi_i
x_i),\nn
\\
\bar Q&=&\int \rmd x
(\rho\upsilon_i\bar\psi_i+\sqrt{2\lambda}\rho^{\frac{\gamma+1}{2}}\bar\sigma-\frac{i}{2}(\gamma-1)\rho\partial_i\bar\psi_i\sigma\bar\sigma-
\frac{i}{R}\rho\bar\psi_i x_i),
\end{eqnarray}
where $\lambda$ is a coupling constant.
They produce the super-extended Hamiltonian via $\{Q,\bar Q\}=2iH$ with
\begin{eqnarray}\label{SupNHHam}
H&=&\int \rmd x
\Big(\frac12\rho\upsilon_i\upsilon_i+\lambda\rho^{\gamma}+\frac{1}{2R^2}
\rho
x_ix_i-i\frac{\sqrt{2\lambda}}{2}(\gamma-1)\rho^\frac{\gamma+1}{2}(\partial_i\psi_i\bar\sigma+\partial_i\bar\psi_i\sigma)\nn\\
&&\qquad\qquad
-\frac12(\gamma-1)\rho\partial_i\psi_j\partial_j\bar\psi_i\sigma\bar\sigma
+\frac{1}{8}(\gamma-1)^2\frac{1}{\rho}\partial_i(\rho\sigma\bar\sigma)\partial_i(\rho\sigma\bar\sigma)\nn
\\
&&\qquad\qquad+\frac{1}{R}\rho\psi_i\bar\psi_i+\frac{d}{2R}(\gamma-1)\rho\sigma\bar\sigma\Big).
\end{eqnarray}
It is also straightforward to verify that $\{Q, Q\}=\{\bar Q,\bar
Q\}=0$. The Hamiltonian differs from (\ref{HamN2}) only by
the contributions with a harmonic potential for both bosonic and
fermionic degrees of freedom.

Equations of motion obtained by bracketing
$\rho,\sigma,\psi_i,\upsilon_i$ with $H$ read
\begin{eqnarray}\label{SupNHeq}
{\cal D}\rho=-\rho\partial_i\upsilon_i,&&\qquad \rho\left({\cal
D}\upsilon_i+\frac{1}{R^2}
x_i\right)=-(\gamma-1)\partial_i(\lambda\rho^{\gamma})+\partial_j\Sigma_{ij}\nn
\\
{\cal
D}\sigma&=&-\frac{id(\gamma-1)}{2R}\sigma-\frac{\sqrt{2\lambda}}{2}(\gamma-1)\rho^{\frac{\gamma-1}{2}}\partial_i\psi_i\nn
\\
&& +i\frac{\gamma-1}{2}\partial_i\psi_j\partial_j\bar\psi_i\sigma
+\frac{i}{4}(\gamma-1)^2\sigma\partial_i(\frac{1}{\rho}\partial_i(\rho\sigma\bar\sigma)),\nn
\\
\rho\left({\cal
D}\psi_i+\frac{i}{R}\psi_i\right)&=&-\frac{\sqrt{2\lambda}}{2}(\gamma-1)\partial_i(\rho^{\frac{\gamma+1}{2}}\sigma)
-i\frac{(\gamma-1)}{2}\partial_j(\rho\partial_i\psi_j\sigma\bar\sigma),
\end{eqnarray}
where $\Sigma_{ij}$ is specified in
(\ref{SigTen}).

After such supersymmetrization, all symmetries of the original
bosonic theory continue to hold but the conserved charges in
(\ref{NHCons}) now must be suitably modified. Namely, while $P_i,
C_i$ and $m$ maintain their form,  $M_{ij},D$ and $K$ in
(\ref{NHCons}) must be modified as follows
\begin{eqnarray}\label{SupNHCons}
M_{ij}&=&\int \rmd x (\rho\upsilon_ix_j-\rho\upsilon_jx_i)-i\int
\rmd x \rho(\psi_i\bar\psi_j-\psi_j\bar\psi_i),\quad J=\int \rmd x
\rho(\psi_i\bar\psi_i+\sigma\bar\sigma)\nn
\\
D&=&\frac{R}{2}(H-\frac{1}{R}J)\sin{\frac{2t}{R}}-\frac{1}{2}\int
dx(\rho\upsilon_ix_i\cos{\frac{2t}{R}}+\frac{1}{R}\rho
x_ix_i\sin{\frac{2t}{R}})\nn \\
K&=&\frac{R^2}{2}(H-\frac{1}{R}J)(1-
\cos{\frac{2t}{R}})-\frac{R}{2}\int \rmd
x(\rho\upsilon_ix_i\sin{\frac{2t}{R}}-\frac{1}{R}\rho
x_ix_i\cos{\frac{2t}{R}}),
\end{eqnarray}
with $H$ in (\ref{SupNHHam}).

In addition to $Q,\bar Q$, in the Grassmann-odd sector of the superalgebra
one has
\begin{align}\label{SupNHConsF}
& \Lambda_i=e^{i\frac{t}{R}}\int \rmd x \rho\psi_i, &&
S=-\frac{i}{2}R(e^{2i\frac{t}{R}}-1)Q-e^{2i\frac{t}{R}}\int \rmd
x\rho\psi_ix_i, \nn
\\
& \bar\Lambda_i=e^{-i\frac{t}{R}}\int \rmd x \rho\bar\psi_i, && \bar
S= \frac{i}{2}R(e^{-2i\frac{t}{R}}-1)\bar Q-e^{-2i\frac{t}{R}}\int
\rmd x\rho\bar\psi_ix_i,
\end{align}
where $(S,\bar S)$ are the generators of the superconformal transformations
and $(\Lambda_i,\bar\Lambda_i)$ are the superpartners of the Galilei boosts.

Calculating the Poisson brackets for the conserved
charges above, one reproduces the $N=2$ conformal Newton-Hooke superalgebra
\cite{Gal09,Gal10}
\begin{align}
&  \{Q,\bar Q\}=2iH, && \{\bar Q,\Lambda_i\}=iP_i-\frac{1}{R}C_i, &&
[H,S]=Q+\frac{2i}{R}S,\nn
\\
& \{Q,\bar S\}=2iD+J+\frac{2}{R}K, && \{S,\bar\Lambda_i\}=iC_i, &&
[D,S]=\frac12S,\nn
\\
& \{S,\bar S\}=2iK, && \{\Lambda_i,\bar\Lambda_j\}=i\delta_{ij}m, &&
[D,Q]=-\frac12Q-\frac{i}{R}S,\nn
\\
& [K,Q]=-S, && [Q,C_i]=\Lambda_i, && [S,P_i]=-\Lambda_i,\nn
\\
& [K,\bar Q]=-\bar S,  && \{ Q,\bar\Lambda_i\}=iP_i+\frac{1}{R}C_i,
&& [H,\bar S]=\bar Q-\frac{2i}{R}\bar S,\nn
\\
& \{\bar Q, S\}=2iD-J-\frac{2}{R}K, && \{\bar S,\Lambda_i\}=iC_i, &&
[D,\bar S]=\frac12\bar S,\nn
\\
& [\bar Q,C_i]=\bar\Lambda_i, && [\bar S,P_i]=-\bar\Lambda_i, &&
[D,\bar Q]=-\frac12\bar Q+\frac{i}{R}\bar S,\nn
\\
& [J,Q]=iQ, && [J,\bar Q]=-i\bar Q,  && [J,S]=iS,\nn
\\
& [J,\bar S]=-i\bar S, && [J,\Lambda_i]=i\Lambda_i, &&
[J,\bar\Lambda_i]=-i\bar\Lambda_i,\nn
\\
& [Q,P_i]=\frac{i}{R}\Lambda_i, && [\bar
Q,P_i]=-\frac{i}{R}\Lambda_i, &&
[H,\Lambda_i]=\frac{i}{R}\Lambda_i\nn
\\
&
[\Lambda_i,M_{jk}^{}]=\delta_{ij}\Lambda_{k}-\delta_{ik}\Lambda_{j},
&&
[\bar\Lambda_i,M_{jk}^{}]=\delta_{ij}\bar\Lambda_{k}-\delta_{ik}\bar\Lambda_{j},
&& [H,\bar\Lambda_i]=-\frac{i}{R}\bar\Lambda_i.
\end{align}
Here we omitted the structure relations for the Grassmann-even part
which read as in (\ref{ShrAl}) and (\ref{NHal}) with the additional
replacement ${[H,D]}=H+\frac{1}{R}J-\frac{2}{{\cal R}^2}K$.

A few comments are in order. Firstly, in the limit of a vanishing
cosmological constant (i.e. $R\rightarrow\infty$) the equations of
motion (\ref{SupNHeq}) as well as the conserved charges
(\ref{SupNHCons}), (\ref{SupNHConsF}) reduce to those of
the perfect fluid dynamics with $N=2$ Schr\"odinger supersymmetry
considered above. On the other hand, in the bosonic limit,
when all the fermionic variables tend to zero, the original
non-supersymmetric theory (\ref{NHHam}) is reproduced.

Secondly, the Clebsch-like decomposition for the velocity vector field
\begin{eqnarray}\label{SupNHCleb}
\upsilon_i=\partial_i\theta+\frac{i}{2}(\psi_j\partial_i\bar\psi_j+\bar\psi_j\partial_i\psi_j)+\frac{i}{2}(\sigma\partial_i\bar\sigma+\bar\sigma\partial_i\sigma)
\end{eqnarray}
can be used to derive the supersymmetric equations of motion
(\ref{SupNHeq}) from the variational principle associated with Lagrangian
\begin{eqnarray}
L&=&-\int \rmd x
\rho\Big(\partial_0\theta+\frac{i}{2}(\psi_j\partial_0\bar\psi_j
+\bar\psi_j\partial_0\psi_j)+\frac{i}{2}(\sigma\partial_0\bar\sigma
+\bar\sigma\partial_0\sigma)\Big)-H.
\end{eqnarray}
Here, $\theta$ is a bosonic prepotential of the velocity and $H$ is
the super-extended Hamiltonian (\ref{SupNHHam}). The variation with
respect to $\theta,\bar\sigma,\bar\psi_i$ gives the continuity
equation and equations of motion for $\sigma,\psi_i$ in
(\ref{SupNHeq}), while varying with respect to $\rho$ and taking
into account (\ref{SupNHCleb}), one obtains the super-extended Euler
equation in (\ref{SupNHeq}). In this formulation fermionic variables
play the role of the Gaussian potentials for $\upsilon_i$ giving
rise to the non-vanishing fluid's vorticity
\begin{eqnarray}
\omega_{ij}=\partial_i\upsilon_j-\partial_j\upsilon_i=
i(\partial_i\psi_k\partial_j\bar\psi_k-\partial_j\psi_k\partial_i\bar\psi_k+\partial_i\sigma\partial_j\bar\sigma-
\partial_j\sigma\partial_i\bar\sigma),
\end{eqnarray}
which in the bosonic limit vanishes as it should be for an irrotational
fluid $\upsilon_i=\partial_i\theta$. A representation of fluid's
vorticity by anticommuting variables was first studied in
\cite{JP00} and recently it was discussed in the fluid
model with $N=1$ Schr\"odinger supersymmetry \cite{Sne25b}.

Thirdly, in the recent paper \cite{Sne25a} it was shown that
non-supersymmetric perfect fluid equations with the Schr\"odinger
symmetry and the conformal Newton-Hooke symmetry are connected by
\begin{eqnarray}\label{FluidNied}
\rho'(t',x')&=&({\cos{\frac{t}{R}}})^{ d}\rho(t,x),\nn
\\
\upsilon'_i(t',x')&=&({\cos{\frac{t}{R}}})\left(\upsilon_i(t,x)+\frac{1}{R}{\tan{\frac{t}{R}}}x_i\right),
\end{eqnarray}
which originate from Niederer's transformation\footnote{The density
transformation is obtained by requiring the mass of a
$d$-dimensional volume element to be invariant
$\int_{V'}dx'\rho'(t',x')=\int_{V}dx\rho(t,x)$, while the
transformation law for $\upsilon_i(t,x)$ is obtained by considering
the orbit of a fluid particle $x_i(t)$ and taking into account that
$\frac{dx_i(t)}{dt}=\upsilon_i(t,x(t))$.}
$t'=R\tan{\frac{t}{R}}$,
$x'_i=(\cos{\frac{t}{R}})^{-1} x_i$ \cite{N73} where coordinates
with prime parameterize the flat space.
The Niederer transformation
(locally) links a free particle to the harmonic oscillator and it
reflects the well-known fact that the Schr\"odinger algebra and the
conformal Newton-Hooke algebra are isomorphic \cite{NOR97}. If one
adds to (\ref{FluidNied}) the transformations of fermions
\begin{align}
& \sigma'(t',x')=e^{i\frac{t}{R}}\sigma(t,x), &&
\psi'_i(t',x')=e^{i\frac{t}{R}}\psi_i(t,x)\nn
\\
& \bar\sigma'(t',x')=e^{-i\frac{t}{R}}\bar\sigma(t,x), &&
\bar\psi'_i(t',x')=e^{-i\frac{t}{R}}\bar\psi_i(t,x)
\end{align}
one observes that the corresponding supersymmetric perfect fluid
equation (\ref{N2eq}) and (\ref{SupNHeq}) are related.

\section{Perfect fluid equations with $\ell$-conformal Galilei\\ supersymmetry}\label{Sec4}

As is well known, the Schr\"odinger algebra is a particular instance
of the so called $\ell$-conformal Galilei algebra
\cite{Hen97,NOR97}. The latter is characterized by an arbitrary
(half)-integer real parameter $\ell$ and it includes generators of time
translations $H$, dilatations $D$, special conformal transformation
$K$, space rotations $M_{ij}$, space translations $C_i^{(0)}$,
Galilei boosts $C_i^{(1)}$, as well as $2\ell-1$
generators of constant accelerations $C_i^{(2)},...,C_i^{(2\ell)}$.
They obey the structure relations
\begin{align}\label{lconfG}
& {[H,D]}=H, && {[H,C^{(k)}_i]}=kC^{(k-1)}_i,\nn
\\
& {[H,K]}=2D, && {[D,C^{(k)}_i]}=(k-\ell)C^{(k)}_i,\nn
\\
& {[D,K]}=K, && {[K,C^{(k)}_i]}={(k-2\ell)}C^{(k+1)}_i,\nn
\\
& {[C_i^{(k)},M_{ab}]}=\delta_{ia}C_{b}^{(k)}-\delta_{ib}C_a^{(k)},
&& {[M_{ij},M_{ab}]}=\delta_{i[a}M_{b]j}-\delta_{j[a}M_{b]i},
\end{align}
where $k=0,1,...,2\ell$ and $i,j,a,b=1,...,d$, $d$ is the spatial
dimension.

In arbitrary dimension and for half-integer $\ell$, the
$\ell$-conformal Galilei algebra admits a central extension
\cite{GM15a}
\begin{eqnarray}\label{CentlconG}
[C_i^{(k)},C_j^{(k')}]&=&(-1)^{k}k!k'!\delta^{(k+k')(2\ell)}\delta_{ij}m,
\end{eqnarray}
where the central charge $m$ links to the mass in dynamical
realizations. For $\ell=\frac12$ the constant accelerations are
absent and (\ref{lconfG}), (\ref{CentlconG}) reduce to the
Schr\"odinger algebra (\ref{ShrAl}), where we identify
$C_i^{(0)}\equiv P_i$ and $C_i^{(1)}\equiv C_i$.

The perfect fluid equations, which accommodate the $\ell$-conformal
Galilei symmetry, were formulated in \cite{Gal24}
\begin{eqnarray}\label{PFEqlCG}
{\partial_0\rho}+ {\partial_i (\rho\upsilon_i)}=0,\quad{\cal
D}^{2\ell}\upsilon_i=-\frac{1}{\rho}{\partial_i p},\quad
p=\nu\rho^{1+\frac{1}{\ell d}}.
\end{eqnarray}
They include the conventual continuity equation for the density $\rho$,
the generalized Euler equation involving $2\ell$ material
derivatives acting upon the velocity $\upsilon_i$, and the polytropic
equation of state with the exponent $\gamma=1+\frac{1}{\ell d}$. For
$\ell=\frac12$ the usual perfect fluid equations with the Schr\"odinger
symmetry are reproduced.

Our goal in this section is to construct supersymmetric extensions
of (\ref{PFEqlCG}). Like before, we choose to work
within the Hamiltonian formalism, which for the model at hand was
constructed in \cite{Sne23a} for half-integer $\ell=n+\frac12$,
$n=0,1,2,...$.

Following \cite{Sne23a}, one needs to introduce
Ostrogratsky-like auxiliary field variables $\upsilon_i^{k}$,
$k=0,1,...,2n$ with $\upsilon_i^{0}=\upsilon_i$ and rewrite the
second equation in (\ref{PFEqlCG}) as the first-order system
\begin{eqnarray}
{\cal D}\upsilon^{k}_i=\upsilon^{k+1}_i,\quad{\cal
D}\upsilon^{2n}_i=-\frac{1}{\rho}{\partial_i p},\quad k=0,1,...2n-1.
\end{eqnarray}
Then the Hamiltonian
\begin{eqnarray}\label{lCGHam}
H&=&\int \rmd x
\left(\frac12\rho\sum_{k=0}^{2n}(-1)^k\upsilon^{k}_i\upsilon^{2n-k}_i+V\right),\quad
V=\lambda\rho^{1+\frac{1}{\ell d}}
\end{eqnarray}
governs the corresponding dynamics, provided the Poisson brackets
\begin{eqnarray}\label{lCGPB}
&&\{\rho(x),\upsilon^{k}_i(x')\}=-\delta^{(k)(2n)}\partial_i\delta(x,x')
\\
\{\upsilon^{k}_i(x),\upsilon^{m}_j(x')\}&=&\frac{1}{\rho}\left(\delta^{(k)(2n)}\partial_i
\upsilon^{m}_j- \delta^{(m)(2n)}\partial_j
\upsilon^{k}_i+(-1)^{k+1}\delta^{(k+m)(2n-1)}\delta_{ij}\right)\delta(x,x')\nn
\end{eqnarray}
are used.

Conserved charges associated with the $\ell$-conformal Galilei
symmetry read
\begin{eqnarray}\label{lCGCons}
D&=&tH-\frac12\int \rmd
x\rho\sum_{k=0}^{2n}(-1)^{k}(k+1)\upsilon_i^{k}\upsilon_i^{2n-k-1},\nn
\\
K&=&-t^2H+2tD+\frac12\int \rmd
x\rho\sum_{k=0}^{2n}(-1)^k\Big((n+1)(2n+1)-k(k+1)\Big)\upsilon^{k-1}_i\upsilon^{2n-k-1}_i,\nn
\\
C_i^{(k)}&=&\sum_{s=0}^k(-1)^s\frac{k!}{(k-s)!}t^{k-s}\int \rmd x
\rho \upsilon^{2n-s}_i,\quad  k=0,...,2n\nn,
\\
M_{ij}&=&\int \rmd x
\rho\sum_{k=0}^{2n+1}(-1)^k\upsilon_i^{2n-k}\upsilon_j^{k-1},
\end{eqnarray}
where the notation $\upsilon^{-1}_i=x_i$ was used. Together with the
Hamiltonian (\ref{lCGHam}), they  form  a  representation  of  the
algebra (\ref{lconfG}), (\ref{CentlconG}) under the Poisson bracket
(\ref{lCGPB}).

\subsection{$N=1$ supersymmetry}

In order to construct the $N=1$ supersymmetric extension for the
dynamical system (\ref{lCGHam}), one introduces real fermionic
fields $\sigma(t,x)$ and $\psi_i^k(t,x)$, $k=0,1,...,2n$ as
superpartners for $\rho$ and
$\upsilon_i^k$. In contrast to the Schr\"odinger case, one has many
fermionic vector variables and it is not evident how to
impose Poisson brackets for them. In this paper, we propose the following
variant
\begin{align}\label{N1lCGPB}
&
\{\psi^{k}_i(x),\psi^{m}_j(x')\}=\frac{i}{\rho}(-1)^{k}\delta^{(k+m)(2n)}\delta_{ij}\delta(x,x'),
&& \{\sigma(x),\sigma(x')\}=\frac{i}{\rho}\delta(x,x'),\nn
\\
&
\{\upsilon_i^{2n}(x),\psi^{k}_j(x')\}=\frac{1}{\rho}\partial_i\psi^{k}_j\delta(x,x'),
&& \{\upsilon^{2n}_i(x),\sigma(x')\}=\frac{1}{\rho}\partial_i
\sigma\delta(x,x').
\end{align}
Taking into account (\ref{lCGPB}), it is straightforward to verify
that the super-Jacobi identities are satisfied.

As the next step, one constructs the real supersymmetry charge
\begin{eqnarray}
Q&=&\int \rmd x
\left(\rho\sum_{k=0}^{2n}(-1)^k\upsilon^{k}_i\psi^{2n-k}_i
+\sqrt{2\lambda}\rho^{\frac{\gamma+1}{2}}\sigma\right),
\end{eqnarray}
which generates the
$N=1$ super-extended Hamiltonian via the Poisson
bracket $\{Q,Q\}=2iH$
\begin{eqnarray}\label{N1lCGHam}
H&=&\int \rmd x
\Big(\frac12\rho\sum_{k=0}^{2n}(-1)^k\upsilon^{k}_i\upsilon^{2n-k}_i+\lambda\rho^\gamma\nn\\
&&\qquad-\frac{i}{2}\rho\sum_{k=1}^{2n}(-1)^k\psi_i^{k}\psi_i^{2n-k+1}
-i\frac{\sqrt{2\lambda}}{2}(\gamma-1)\rho^{\frac{\gamma+1}{2}}\partial_i
\psi_i^0\sigma\Big),
\end{eqnarray}
where the terms entering the first line correspond to the original bosonic theory
(\ref{lCGHam}). The fermionic
contributions in second line include the kinetic term for $\psi_i^k$
and the boson-fermion
coupling.

The Hamiltonian (\ref{N1lCGHam}) gives rise to the supersymmetric equations
of motion
\begin{align}\label{N1lCGeq}
& \partial_0\rho+\partial_i(\rho\upsilon_i^0)=0 && {\cal
D}\sigma=-\frac{\sqrt{2\lambda}}{2}(\gamma-1)\rho^{\frac{\gamma-1}{2}}\partial_i\psi_i^0\nn
\\
&{\cal D}\upsilon_i^k=\upsilon_i^{k+1} && {\cal
D}\psi_i^k=\psi_i^{k+1}\nn
\\
&\rho{\cal
D}\upsilon_i^{2n}=-\lambda(\gamma-1)\partial_i(\rho^{\gamma})+\partial_j\Sigma_{ij}
&& \rho{\cal
D}\psi_i^{2n}=-\frac{\sqrt{2\lambda}}{2}(\gamma-1)
\partial_i(\rho^{\frac{\gamma+1}{2}}\sigma),
\end{align}
where
$$
\Sigma_{ij}=i\frac{\sqrt{2\lambda}}{4}(\gamma-1)\Big((\gamma-1)\delta_{ij}\partial_k\psi^0_k
+2\partial_i\psi^0_j\Big)\rho^{\frac{\gamma+1}{2}}\sigma.
$$
One can see that both the bosonic and  fermionic vector fields
$\upsilon_i^k$ and $\psi_i^k$, $k\geq1$ are auxiliary. They can be
found from (\ref{N1lCGeq}) as higher-order material
derivatives of the master fields $\upsilon_i^k={\cal
D}^k\upsilon_i^0$ and $\psi_i^k={\cal D}^k\psi_i^0$.

Conserved charges associated with symmetries of the original bosonic
theory read
\begin{eqnarray}\label{N1lCGConsB}
D&=&tH-\frac12\int \rmd
x\rho\sum_{k=0}^{2n}(-1)^{k}(k+1)\upsilon_i^{k}\upsilon_i^{2n-k-1}-\frac{i}{2}\int
\rmd x\rho \sum_{k=0}^{2n}(-1)^k(n-k)\psi_i^{k}\psi_i^{2n-k},\nn
\\
K&=&-t^2H+2tD+\frac12\int \rmd
x\rho\sum_{k=0}^{2n}(-1)^k\Big((n+1)(2n+1)-k(k+1)\Big)\upsilon^{k-1}_i\upsilon^{2n-k-1}_i\nn
\\
&&-\frac{i}{2}\int \rmd
x\rho\sum_{k=1}^{2n}(-1)^{k}k(2n-k+1)\psi_i^{k-1}\psi_i^{2n-k},\nn
\\
C_i^{(k)}&=&\sum_{s=0}^k(-1)^s\frac{k!}{(k-s)!}t^{k-s}\int \rmd x
\rho \upsilon^{2n-s}_i,\quad  k=0,...,2n+1,\nn
\\
M_{ij}&=&\int \rmd x
\rho\sum_{k=0}^{2n+1}(-1)^k\upsilon_i^{2n-k}\upsilon_j^{k-1}-i\int
\rmd x\rho \sum_{k=0}^{2n}(-1)^k\psi_i^{2n-k}\psi_j^{k}.
\end{eqnarray}
Taking into account (\ref{lCGCons}), one concludes that the supersymmetric
procedure does not alter the form of $C_i^{(k)}$.

Conserved charges in the fermionic sector read
\begin{eqnarray}\label{N1lCGConsF}
S&=&tQ-\int \rmd x
\rho\sum_{k=0}^{2n}(-1)^k(2n-k+1)\upsilon^{k-1}_i\psi^{2n-k}_i\nn
\\
\Lambda_i^{(k)}&=&\sum_{s=0}^k(-1)^s\frac{k!}{(k-s)!}t^{k-s}\int
\rmd x \rho \psi^{2n-s}_i,\quad  k=0,...,2n.
\end{eqnarray}

Under the Poisson bracket,  (\ref{N1lCGConsB})
and (\ref{N1lCGConsF}) form the $N=1$ $\ell$-conformal Galilei
superalgebra \cite{Aiz12,GM21}
\begin{align}
&  \{Q,Q\}=2iH, &&  [H,\Lambda_i^{(k)}]=k\Lambda_i^{(k-1)} ,\nn
\\
& \{Q,S\}=2iD, && [D,\Lambda_i^{(k)}]=(k-n)\Lambda_i^{(k)} ,\nn
\\
& \{S,S\}=2iK, && [K,\Lambda_i^{(k)}]=(k-2n)\Lambda_i^{(k+1)} ,\nn
\\
& [K,Q]=-S, && \{Q,\Lambda_i^{(k)}\}=iC_i^{(k)} \nn
\\
& [H,S]=Q && \{S,\Lambda_i^{(k)}\}=iC_i^{(k+1)}\nn
\\
& [D,S]=\frac12S && \{Q,C_i^{(k)}\}=k\Lambda_i^{(k-1)}\nn
\\
& [D,Q]=-\frac12Q && \{S,C_i^{(k)}\}=(k-2n-1)\Lambda_i^{(k)}\nn
\\
&
[\Lambda_i^{(k)},M_{jk}^{}]=\delta_{ij}\Lambda_{k}^{(k)}-\delta_{ik}\Lambda_{j}^{(k)},
&&
\{\Lambda_i^{(k)},\Lambda_j^{(k')}\}=i(-1)^kk!k'!\delta^{k+k',2n}\delta_{ij}m
\end{align}
It reduces to the $N=1$ Schr\"odinger superalgebra for
$\ell=\frac12$ ($n=0$).

Concluding this section, let us discuss a Lagrangian formulation for
the perfect fluid equations with the $\ell$-conformal Galilei supersymmetry
(\ref{N1lCGeq}). For the non-supersymmetric case, it was
constructed in \cite{Sne24} and it was based on a generalization of the Clebsh
decomposition for the vector variables $\upsilon_i^k$, $k=0,..,2n$. In
particular, it was shown that in order to obtain the equation from
the variational principle only the highest component $\upsilon_i^{2n}$
should be Clebsh-decomposed, while the remaining vector
variables with
$k<2n$ remain intact. Adapting this idea to the supersymmetric case, we choose a
suitable Clebsch-type decomposition for $\upsilon_i^{2n}$ in the
form
\begin{eqnarray}\label{SuplCGCleb}
\upsilon^{2n}_i=\partial_i\theta
+\frac12\sum_{k=0}^{2n-1}(-1)^{k+1}\upsilon_j^{k}\partial_i\upsilon_j^{2n-k-1}
+\frac{i}{2}\sum_{k=0}^{n}(-1)^{k}\psi_j^{k}\partial_i\psi_j^{2n-k}
+\frac{i}{2}\sigma\partial_i\sigma.
\end{eqnarray}
Then the supersymmetric Lagrangian reads
\begin{eqnarray}
L&=&-\int \rmd x
\rho\left(\partial_0\theta+\frac12\sum_{k=0}^{2n-1}(-1)^{k+1}\upsilon_i^{(k)}\partial_0\upsilon_i^{(2n-k-1)}\right.\nn
\\
&&\qquad\qquad
\left.+\frac{i}{2}\sum_{k=0}^{2n}(-1)^{k}\psi_j^{k}\partial_0\psi_j^{2n-k}+\frac{i}{2}\sigma\partial_0\sigma\right)
-H ,
\end{eqnarray}
where $H$ is the Hamiltonian in (\ref{N1lCGHam}) and $\theta$ is the
bosonic prepotential. By varying the Lagrangian with respect to
$\theta$ and $\sigma$, one obtains the the first line in
(\ref{N1lCGeq}). Varying with respect to $\upsilon_i^k$ and
$\upsilon_i^k$ and taking into account the continuity equation, the
second line in (\ref{N1lCGeq}) is reproduced. Finally, varying
with respect to $\rho$, one gets
\begin{eqnarray}
{\cal
D}\theta-\frac12\upsilon^{0}_i\upsilon^{2n}_i+V'_\rho-\frac{i}{4}\sqrt{2\lambda}(\gamma-1)\Big(\gamma\rho^{\frac{\gamma-1}{2}}\partial_i\psi^0_i\sigma
+\frac{1}{\rho}\psi^0_i\partial_i(\rho^{\frac{\gamma+1}{2}}\sigma)\Big)=0.
\end{eqnarray}
As a consequence, the equation of motion  for $\upsilon_i^{2n}$ in
(\ref{N1lCGeq}) is satisfied as well by virtue of
(\ref{SuplCGCleb}).

\subsection{On $N=2$ supersymmetry}

In this section, we discuss an attempt to construct an
$N=2$ supersymmetric extension of
the perfect fluid dynamics with the $\ell$-conformal Galilei symmetries. For
$\ell=\frac12$, the corresponding case was studied in Sect.
\ref{Sec2} and leads to the perfect fluid equations with the $N=2$
Schr\"odinger supersymmetries. As shown below, for $\ell>\frac12$
the situation alters drastically.

Like before, in order to realize the $N=2$ supersymmetry, we introduce
complex fermionic superpartners which for the model at hand include
$\sigma(t,x)$ and $\psi_i^k(t,x)$, $k=0,1,...,2n$ with
complex-conjugates $\bar\sigma(t,x)$ and $\bar\psi_i^k(t,x)$.
A comparison with (\ref{N1lCGPB}) suggests imposing the
Poisson brackets
\begin{align}\label{N2lCGPB}
&
\{\psi^{k}_i(x),\bar\psi^{m}_j(x')\}=\frac{i}{\rho}(-1)^{k}\delta^{(k+m)(2n)}\delta_{ij}\delta(x,x')
&& \{\sigma(x),\bar\sigma(x')\}=\frac{i}{\rho}\delta(x,x')\nn
\\
&
\{\upsilon_i^{2n}(x),\psi^{k}_j(x')\}=\frac{1}{\rho}\partial_i\psi^{k}_j\delta(x,x')
&& \{\upsilon^{2n}_i(x),\sigma(x')\}=\frac{1}{\rho}\partial_i
\sigma\delta(x,x')\nn
\\
&
\{\upsilon_i^{2n}(x),\bar\psi^{k}_j(x')\}=\frac{1}{\rho}\partial_i\bar\psi^{k}_j\delta(x,x')
&& \{\upsilon^{2n}_i(x),\bar\sigma(x')\}=\frac{1}{\rho}\partial_i
\bar\sigma\delta(x,x').
\end{align}

At the next step, one constructs two complex conjugate supercharges
$Q$ and $\bar Q$ which satisfy the relations
\begin{eqnarray}\label{SupRel}
\{Q,Q\}=0,\quad \{\bar Q,\bar Q\}=0,\quad \{Q,\bar Q\}=2iH,
\end{eqnarray}
under the Poisson bracket (\ref{lCGPB}), (\ref{N2lCGPB}).
Here, the superextended Hamiltonian $H$ reduces to the initial bosonic
Hamiltonian (\ref{lCGHam}) in the limit
of vanishing fermionic variables. In general, in order to construct a
supercharge, one has to consider a polynomial
expansion
\begin{eqnarray}
Q=Q_1+Q_3+Q_5+...
\end{eqnarray}
where $Q_1,Q_3,Q_5,...$ denote linear, cubic, fifth-order etc.
contributions in the fermionic fields. For the case at hand, it proves sufficient
to introduce
only the cubic terms
\begin{eqnarray}
Q&=&\int \rmd x
\Big(\rho\sum_{k=0}^{2n}(-1)^k\upsilon^{k}_i\psi^{2n-k}_i+\sqrt{2\lambda}\rho^{\frac{\gamma+1}{2}}\sigma\nn\\
&&+\frac{i}{2}(\gamma-1)\rho\partial_i\psi_i^0\sigma\bar\sigma
+\frac{i}{2\sqrt{2\lambda}}\rho^{\frac{3-\gamma}{2}}\sum_{k=1}^{2n}(-1)^k\psi_i^{k}\psi_i^{2n-k+1}\bar\sigma\Big),\nn
\\
\bar Q&=&\int \rmd x
\Big(\rho\sum_{k=0}^{2n}(-1)^k\upsilon^{k}_i\bar\psi^{2n-k}_i+\sqrt{2\lambda}\rho^{\frac{\gamma+1}{2}}\bar\sigma\nn
\\
&&-\frac{i}{2}(\gamma-1)\rho\partial_i\bar\psi_i^0\sigma\bar\sigma
+\frac{i}{2\sqrt{2\lambda}}\rho^{\frac{3-\gamma}{2}}\sum_{k=1}^{2n}
(-1)^k\bar\psi_i^{k}\bar\psi_i^{2n-k+1}\sigma\Big),
\end{eqnarray}
where $\lambda$ is a coupling constant.
Interestingly, in contrast to the $N=2$ supercharges in the Schr\"odinger
case $\ell=\frac12$ ($n=0$), the cubic contribution involves an
additional term which contains coupling constant $\lambda$ in
the denominator such that the limit $\lambda\rightarrow0$ is singular. In
particular, it means that for $\ell>\frac12$ the above construction
cannot be implemented in the "free" case $\lambda=0$.

For simplicity of the presentation, in what follows
we focus on  $\ell=\frac32$ which is obtained from the above formulas by
setting $n=1$. In this case, the supersymmetric fluid model is described
by the bosons $\rho,\upsilon_i^0,\upsilon_i^1,\upsilon_i^2$ and the fermions
$\rho,\psi_i^0,\psi_i^1,\psi_i^2$, while their dynamics is governed by
the Hamiltonian which follows from (\ref{SupRel})
\begin{eqnarray}\label{N2lCGHam}
H&=&\int \rmd x
\Big(\rho\upsilon^{2}_i\upsilon^{0}_i-\frac12\rho\upsilon^{1}_i\upsilon^{1}_i+\lambda\rho^\gamma\nn\\
&&-\frac{i}{2}\rho(\psi_i^2\bar\psi_i^1+\bar\psi_i^2\psi_i^1)
-\frac{i}{2}\sqrt{2\lambda}(\gamma-1)\rho^{\frac{\gamma+1}{2}}(\partial_i\psi_i^0\bar\sigma+\partial_i\bar\psi_i^0\sigma)\nn\\
&&-\frac{i}{2\sqrt{2\lambda}}\rho^{\frac{3-\gamma}{2}}\upsilon_i^1(\bar\psi_i^2\sigma+\psi_i^2\bar\sigma)
+\frac{i}{2\sqrt{2\lambda}}\rho^{\frac{3-\gamma}{2}}\upsilon_i^2(\bar\psi_i^1\sigma+\psi_i^1\bar\sigma)\nn
\\
&&+\frac{1}{4\sqrt{2\lambda}}(\gamma-1)\rho^{\frac{1-\gamma}{2}}\partial_i(\rho\sigma\bar\sigma)(\bar\psi_i^1\sigma-\psi_i^1\bar\sigma)\nn
\\
&&+\frac{1}{4\lambda}\rho^{2-\gamma}(\psi_i^2\bar\psi_i^2\sigma\bar\sigma+\psi_i^2\bar\psi_j^2\psi_i^1\bar\psi_j^1)
-\frac12(\gamma-1)\rho\partial_i\psi_j^0\partial_j\bar\psi_i^0\sigma\bar\sigma\nn
\\
&&-\frac{1}{2\sqrt{2\lambda}}(\gamma-1)\rho^{\frac{3-\gamma}{2}}(\partial_i\psi_i^0\bar\psi_j^2\bar\psi_j^1\sigma+
\partial_i\bar\psi_i^0\psi_j^2\psi_j^1\bar\sigma)\Big).
\end{eqnarray}

Turning to symmetries of the model, $H$ and $Q$ are conserved due to
$\{H,H\}=0$, $\{Q,H\}=0$, and
$\{\bar Q,H\}=0$.  They link to the invariance under time translations
and supersymmetry transformations. In order to construct conserved charges
associated
with other symmetries, it is convenient to use the explicit form of
the equations of motion (see Appendix). For example, taking into
account (\ref{lCGPB}), the equation of motion for the density
$\partial_0\rho=\{\rho,H\}$ can be brought to the form
\begin{eqnarray*}
\partial_0\rho&+&\partial_i(\rho\upsilon_i^0+
\frac{i}{2\sqrt{2\lambda}}\rho^{\frac{3-\gamma}{2}}
(\bar\psi_i^1\sigma+\psi_i^1\bar\sigma))=0.
\end{eqnarray*}
Likewise, it
is straightforward to verify that under the Poisson bracket
the following conserved charges
\begin{eqnarray}
C_i^0&=&\int \rmd x\rho\upsilon_i^2,\qquad C_i^1=tC_i^0-\int \rmd
x\rho\upsilon_i^1\nn
\\
M_{ij}&=&\int \rmd
x\rho(\upsilon^2_ix_j-\upsilon^2_jx_i+\upsilon^0_i\upsilon^1_j-
\upsilon^0_j\upsilon^1_i)-i\int \rmd
x(\psi_{[i}^0\bar\psi_{j]}^2+\bar\psi_{[i}^0\psi_{j]}^2-\psi_{[i}^1\bar\psi_{j]}^1)\nn
\\
\Lambda_i^0&=&\int \rmd x\rho\psi_i^2!,\qquad J=\int \rmd
x\rho(\psi_{i}^0\bar\psi_{i}^2-\bar\psi_{i}^0\psi_{i}^2-\psi_{i}^1\bar\psi_{i}^1+\sigma\bar\sigma)
\end{eqnarray}
together with $H,Q,\bar Q$ constructed above, form the $N=2$
supersymmetric extension of the Galilei algebra with the central
charge $m=\int\rmd x\rho$. Here the bosonic generators
$C_i^0,C_i^1,M_{ij},$ and $J$ correspond to space translation,
Galilei boost, space rotations and $u(1)$ $R$-symmetry, while the
fermionic generator $\Lambda_i^0$ provides a superpartner to the
Galilei boost.

Surprisingly enough, our attempts to build
conserved charges associated with the
dilatation, special conformal transformation, accelerations, and
superconformal transformation failed. A possible way to avoid this
difficulty is to consider more involved
fluid system. As was mentioned in the Introduction, the $N=2$
$\ell$-conformal Galilei superalgebra \cite{GM21} must necessarily
include extra bosonic generators for $\ell>\frac12$. This indicates
that the corresponding dynamical realizations may require extra bosonic degrees of freedom. Exactly this way
was used to construct a dynamical system with the $N=2$
$\ell$-conformal Galilei supersymmetry in classical mechanics
\cite{Mas14}.  It is plausible that similar extra bosonic fluid variables must
be introduced in order to realize the desired symmetries. This possibility will be
explored elsewhere.

\section{Conclusion}\label{Sec5}

To summarize, in this work our recent result on the construction
of the perfect fluid equations with the $N=1,2$ Schr\"odinger supersymmetry
\cite{Sne25b} was generalized to other cases of interest. In particular,
the generalized
perfect fluid equations with $N=1$ $\ell$-confomal Galilei
supersymmetry or the
$N=2$ conformal Newton-Hooke supersymmetry were built.
For both the cases, it proved sufficient to introduce the conventual
superpartners for fluid variables and build
 the
supersymmetry charges, which generated the super-extended Hamiltonian
via the Poisson bracket. The latter governed the dynamics of the
resulting system.  The full set of conserved charges associated with
the superalgebras was built. The Clebsh-like decomposition for the
fluid velocity vector and its generalization was used to formulate
the Lagrangian descriptions.

Somewhat unexpectedly, a similar consideration
of perfect fluid
dynamics with the $N=2$ $\ell$-conformal Galilei supersymmetry  failed
for $\ell>\frac12$. Although it proved straightforward to construct an
$N=2$ supersymmetric extension of the Galilei subalgebra, an extension of
the conformal symmetries
of the original bosonic theory proved problematic.
A possibility to introduce extra
bosonic variables in order to cure the problem is an interesting
point for future research.

Let us discuss directions in which the present work can be extended.
The paper was
mostly focused on the mathematical structure of fluid mechanics with
the nonrelativistic conformal supersymmetries. It would be
interesting to understand whether they may prove useful in actual physical
contexts. Exact solutions for the superextended fluid systems
reported in Sec. \ref{Sec3} and Sec. \ref{Sec4} are worth studying as well.
In this regard, it is interesting to explore whether the solution generating
techniques in
\cite{Gal26,GH13} can be applied for the models under consideration.
The construction of fluid models
with the $N=3,4$ nonrelativistic conformal supersymmetry is a
problem deserving of study. An extension of the present analysis to the
fluid models with
the Lifshitz symmetry \cite{Gal22b} is worth exploring as well.


\section*{Acknowledgements}
This work was supported by the Russian Science Foundation, grant No
23-11-00002-Ext.

\section*{Appendix: Generalized $N=2$ supersymmetric fluid equation}
In this Appendix, we display the Hamiltonian equations of motion for the
bosonic fields
$\rho,\upsilon_i^0,\upsilon_i^1,\upsilon_i^2$ and the fermionic fields
$\rho,\psi_i^0,\psi_i^1,\psi_i^2$, which follow
from the super-extended Hamiltonian (\ref{N2lCGHam}) and the  Poisson brackets
(\ref{lCGPB}), (\ref{N2lCGPB}) for the case $\ell=\frac32$ ($n=1$)
\begin{eqnarray*}
\partial_0\rho&+&\partial_i(\rho\upsilon_i^0+\frac{i}{2\sqrt{2\lambda}}\rho^{\frac{3-\gamma}{2}}(\bar\psi_i^1\sigma+\psi_i^1\bar\sigma))=0,
\\
{\mathfrak D}\upsilon_i^0&=&\upsilon_i^1
+\frac{i}{2\sqrt{2\lambda}}\rho^{\frac{1-\gamma}{2}}(\bar\psi_i^2\sigma+\psi_i^2\bar\sigma),
\\
{\mathfrak D}\upsilon_i^1&=&\upsilon_i^2,
\\
{\mathfrak D}\upsilon_i^2&=&
-(\gamma-1)\lambda\frac{1}{\rho}\partial_i(\rho^\gamma)+\frac{1}{\rho}\partial_j\Sigma_{ij},
\\
{\mathfrak D}\sigma&=&
-\frac12\sqrt{2\lambda}(\gamma-1)\rho^{\frac{\gamma-1}{2}}\partial_i\psi_i^0-\frac{1}{2\sqrt{2\lambda}}\rho^{\frac{1-\gamma}{2}}\upsilon_i^1\psi_i^2
+\frac{1}{2\sqrt{2\lambda}}\rho^{\frac{1-\gamma}{2}}\upsilon_i^2\psi_i^1
\\
&&\qquad\qquad
+\frac{i}{4\sqrt{2\lambda}}(\gamma-1)\sigma\partial_i(\rho^{\frac{1-\gamma}{2}}(\bar\psi_i^1\sigma-\psi_i^1\bar\sigma))
+\frac{i}{4\sqrt{2\lambda}}(\gamma-1)\rho^{\frac{-1-\gamma}{2}}\psi_i^1\partial_i(\rho\sigma\bar\sigma)
\\
&&\qquad\qquad-\frac{i}{4\lambda}\rho^{1-\gamma}\psi_i^2\bar\psi_i^2\sigma+
\frac{i}{2}(\gamma-1)\partial_i\psi_j^0\partial_j\bar\psi_i^0\sigma
+\frac{i}{2\sqrt{2\lambda}}(\gamma-1)\rho^{\frac{1-\gamma}{2}}\partial_i\bar\psi_i^0\psi_j^2\psi_j^1,
\\
{\mathfrak D}\psi_i^0&=&\frac12\psi_i^1
+\frac{1}{2\sqrt{2\lambda}}\rho^{\frac{1-\gamma}{2}}\upsilon_i^1\sigma-\frac{i}{4\lambda}\rho^{1-\gamma}(\psi_i^2\sigma\bar\sigma+\psi_j^2\psi_j^1\bar\psi_i^1)
\\
&&\qquad\qquad+\frac{i}{2\sqrt{2\lambda}}(\gamma-1)\rho^{\frac{1-\gamma}{2}}\partial_k\psi_k^0\bar\psi_i^1\sigma,
\\
{\mathfrak D}\psi_i^1&=&\frac12\psi_i^2
+\frac{1}{2\sqrt{2\lambda}}\rho^{\frac{1-\gamma}{2}}\upsilon_i^2\sigma+\frac{i}{4\lambda}\rho^{1-\gamma}\psi_j^1\psi_j^2\bar\psi_i^2
\\
&&\qquad\qquad-\frac{i}{4\sqrt{2\lambda}}(\gamma-1)\rho^{-\frac{1+\gamma}{2}}\partial_i(\rho\sigma\bar\sigma)\sigma
+\frac{i}{2\sqrt{2\lambda}}(\gamma-1)\rho^{\frac{1-\gamma}{2}}\partial_k\psi_k^0\bar\psi_i^2\sigma,
\\
{\mathfrak D}\psi_i^2&=&
-\frac{\sqrt{2\lambda}}{2}(\gamma-1)\frac{1}{\rho}\partial_i(\rho^{\frac{\gamma+1}{2}}\sigma)
-\frac{i}{2}(\gamma-1)\frac{1}{\rho}\partial_j(\rho\partial_i\psi_j^0\sigma\bar\sigma)
\\
&&\qquad\qquad+\frac{i}{2\sqrt{2\lambda}}(\gamma-1)\frac{1}{\rho}\partial_i(\rho^{\frac{3-\gamma}{2}}\psi_j^2\psi_j^1\bar\sigma),
\end{eqnarray*}
where $ {\mathfrak
D}=\partial_0+(\upsilon_i^0+\frac{i}{2\sqrt{2\lambda}}\rho^{\frac{1-\gamma}{2}}(\bar\psi_i^1\sigma+\psi_i^1\bar\sigma))\partial_i
$ is the modified material derivative and
\begin{eqnarray*}
\Sigma_{ij}&=&(\gamma-1)\Big[-\frac{i}{2}\sqrt{2\lambda}[
-\frac12(\gamma-1)\delta_{ij}(\rho^{\frac{\gamma+1}{2}}(\partial_k\psi_k^0\bar\sigma+\partial_j\bar\psi_j^0\sigma)
-(\rho^{\frac{\gamma+1}{2}}(\partial_i\psi_j^0\bar\sigma+\partial_i\bar\psi_j^0\sigma)]
\\
&&-\frac{i}{4\sqrt{2\lambda}}\delta_{ij}\rho^{\frac{3-\gamma}{2}}\upsilon_k^1(\bar\psi_k^2\sigma+\psi_k^2\bar\sigma)
+\frac{i}{4\sqrt{2\lambda}}\delta_{ij}(\rho^{\frac{3-\gamma}{2}}\upsilon_k^2(\bar\psi_k^1\sigma+\psi_k^1\bar\sigma))
\\
&&+\frac{1}{4\sqrt{2\lambda}}[
\frac12(\gamma-1)\delta_{ij}\rho^{\frac{1-\gamma}{2}}\partial_k(\rho\sigma\bar\sigma)(\bar\psi_k^1\sigma-\psi_k^1\bar\sigma)
+\rho\sigma\bar\sigma\partial_i(\rho^{\frac{1-\gamma}{2}}(\bar\psi_j^1\sigma-\psi_j^1\bar\sigma))]
\\
&&+\frac{1}{4\lambda}\delta_{ij}\rho^{2-\gamma}(\psi_k^2\bar\psi_k^2\sigma\bar\sigma
+\psi_k^2\bar\psi_l^2\psi_k^1\bar\psi_l^1)
-\frac12(-\partial_i\psi_k^0\partial_k\bar\psi_j^0
+\partial_i\bar\psi_k^0\partial_k\psi_j^0)\rho\sigma\bar\sigma
\\
&&
-\frac{1}{2\sqrt{2\lambda}}[\frac12(\gamma-1)\delta_{ij}(\rho^{\frac{3-\gamma}{2}}\partial_l\psi_l^0\bar\psi_k^2\bar\psi_k^1\sigma+
\rho^{\frac{3-\gamma}{2}}\partial_l\bar\psi_l^0\psi_k^2\psi_k^1\bar\sigma)\\
&&-(\rho^{\frac{3-\gamma}{2}}\partial_i\psi_j^0\bar\psi_k^2\bar\psi_k^1\sigma+
\rho^{\frac{3-\gamma}{2}}\partial_i\bar\psi_j^0\psi_k^2\psi_k^1\bar\sigma)]\Big].
\end{eqnarray*}

\end{document}